**Title.**

Rapid Virtual Simulations: Achieving "Satisficing Learning Impact" with "Realistic-Enough" Activities in Health Science Education

**Authors.**

**Emmanuel G. Blanchard, Ph.D.**

Associate Professor, Laboratoire d'Informatique de l'Université du Mans, Le Mans Université, FRANCE

Adjunct Professor, Department of Surgery, McGill University, CANADA

https://orcid.org/0009-0006-0454-5739

**Jeffrey Wiseman, M.D., F.R.C.P.C.**

Assistant Professor, Faculty of Medicine and Health Sciences, McGill University, CANADA

Faculty, Institute for Health Science Education, McGill University, CANADA

Director of Education, Steinberg Centre for Simulation and Interactive Learning, McGill University, CANADA

**Corresponding Author.**

**Emmanuel G. Blanchard**

Laboratoire d'informatique de l'université du Mans, Avenue Olivier Messiaen

emmanuel.blanchard@univ-lemans.fr

**Rapid Virtual Simulations: Achieving "Satisficing Learning Impact" with "Realistic-Enough" Activities in Health Science Education**

Simulation is an increasingly important part of Health Science Education[1,2]. It fosters active participation and a sense of immersion in authentic clinical activities followed by debriefing or feedback conversations[3] with the aim of making practice in real clinical settings safer[4]. Historically, simulation-based education was largely an in-person activity[5]. However, this approach poses several types of challenges:

- **Human challenges.** Human administrators, actors, and health sciences professionals are needed to create, run and debrief simulation scenarios[6].
- **Location challenges.** A physical space must be reserved for the activity to take place, and health professional learners and educators must leave their busy practices to travel to these spaces[7].
- **Equipment challenges.** Equipment and other technologies may have to be gathered and reserved for the time of the activity[6].
- **Planning and logistical challenges.** One must ensure that all the above elements are available and ready in time for the activity to run smoothly[6]
- **Adaptability challenges.** It is extremely difficult to change a planned in-person educational activity in response to sudden unpredictable events like the COVID pandemic, electrical outages and bad weather precluding easy travel[8].
- **Financial challenges.** Each of the prior challenges impose costs, making in-person simulations expensive[9,10].

Digital technologies can readily virtualize many in-person educational activities, i.e. transform them into virtual environments within which humans can interact with other humans and/or artificial agents[11]. Virtual simulations are associated with large educational effect sizes when

compared to no intervention[12,13], and have several notable advantages to their in-person counterparts:

- **Simplified logistics.** Most elements required for a simulation activity can be recreated virtually. Once virtualized, space or equipment availability is no longer an issue. Similarly, replacing human participants (i.e., actors or health professionals) by virtual pedagogical agents[14] can reduce recruitment and training challenges. Overall, virtualizations of participants, space, and equipment can decrease planning and logistic challenges[15].
- **Repeatability and scalability.** Virtual simulations can be repeated endlessly to support spaced learning[16,17], and/or offered to large numbers of learners without requiring a drastic increase of dedicated resources[18,19].
- **Spatial independence.** Virtual simulations can be used from anywhere, i.e., without location constraints[20].
- **Temporal independence.** Similarly, most virtual simulations can be used at any time, i.e., without time constraints[20].
- **Data-driven smart assessment.** Recording learners' interactions within a virtual simulation is relatively easy. Visualization techniques or AI algorithms can then use such data to identify learners' cognitive, motivational, or affective patterns to, for instance, improve a subsequent debriefing session.

However, virtual simulations also have disadvantages compared to in-person simulations:

- **Immersion perception constraints.** A sense of immersion is much harder to achieve for many learning tasks[21].
- **Interdisciplinary expertise requirements.** Developing such simulations requires additional skills such as computer programming, or virtual content design and editing (2D image, 3D video, sound, video…), making interdisciplinary collaboration mandatory and costly while slowing down project maturation. However, as with many educational

- technology projects, investing more money or expert time does not always guarantee more effective educational experiences for learners[22].
- **Techno-pedagogical challenges.** Virtual simulations require literacy in both technology and education, which raises development complexity. For instance, creating virtual pedagogical agents (i.e. artificial entities capable of reacting in a realistic way to unexpected learner behaviors or requests) is a notoriously complex task[23]. Additionally educators using virtual simulations need to adapt or learn techno-pedagogical skills in order to create educationally effective virtual simulations[24,25].

The concept of *Rapid Virtual Simulations* (RVS) has emerged from a multi-year interdisciplinary health sciences education project as a new class of techno-pedagogical tool that maximizes the benefits and minimizes the disadvantages of virtual simulations. The goal of this paper is to present an argument for RVS as technology-supported interactive learning experiences that are quick to create, deploy, and modify.

This remainder of this paper is organized as follows. The next section describes the key theoretical concepts on which RVS are grounded. Then we provide general definitions for an RVS and an RVS ecosystem, and a suggested modular architecture to implement them with illustrations stemming from a project for Internal Medicine Education. Major benefits and challenges specific to RVS are then presented before discussing reasons for and against their adoption in educational settings. Finally, in conclusion, we open the conversation on future possible extensions of this new techno-pedagogical practice.

**KEY THEORETICAL CONCEPTS**

As mentioned earlier, many virtual simulation projects dedicate a lot of resources towards the creation and editing of digital contents and animations of very high quality. While this approach has advantages such as an increased realism, it makes designers of educational activities dependent on outsiders (e.g. IT experts, multimedia artists), whose production requires validation before pedagogical use. This significantly increases complexity, time, cost, and expenses for case creation and improvement processes. To address this, we propose two crucial and interrelated guiding concepts for the development of RVS: "Satisficing Learning Impact" and "Realistic-Enough" activities .

**Satisficing Learning Impact**

Satisficing[26,27] is a heuristic approach defined as choosing an acceptable or "good-enough"[28] result rather than searching for the perfect result. This permits adaptation to a real environment under time pressure and with limited information and resources Consequently, a Satisficing Learning Impact is defined as a learning impact i.e. changes in a learner's state (alterations in attitudes, skills, knowledge, and emotions resulting from a learning activity) that educators judge to be "satisficing" i.e. acceptably close enough to planned learning objectives.

**Realistic-Enough Activity**

In the context of this paper, we define a Realistic-Enough Activity as a learning activity whose degree of realism is deemed appropriate to the achievement of a satisficing learning impact. This is closely related to Kneebone's use of the same term[29] even though we developed our initiative independently.

However, the concept of realism in healthcare simulation is hotly debated. It is defined by some as the degree to which a simulation reproduces all that occurs in a real situation while others include terms such as fidelity and meaningfulness [8,30–32]. Striving to achieve the highest possible degree of realism is not only costly and time consuming, it is also unnecessary for many learning objectives such as for medical students learning to use an algorithm to prioritize patient care options[33], or mid-level general surgery residents learning critical resource management skills in an operating room[34]. In contrast, others define realism as the degree of *functional task alignment*, i.e. the degree of match between a simulated task, a targeted learning task, and intended learning objectives[35]. Such a perspective on realism permits simulation designers to focus resources and time on creating this functional task alignment rather than on bells and whistles that do not contribute to learning. This second interpretation of realism is the one we have endorsed in the context of RVS and, as a follow up, we have identified two dimensions to manipulate in order to alter realism while seeking functional task alignment:

- **The Domain Model Dimension.** A domain model refers to a simplification of a field of knowledge. For instance, for an internal medicine simulation, the domain model would model concepts, practices and strategies that are the most relevant to specific internal medicine tasks and cases. However, details related to other internal medicine surgical, pediatric, and psychiatric cases do not necessarily have to be included. When designing an RVS, one should create the most simplified knowledge model needed for a pedagogical simulation that ensures functional task alignment towards a satisfying learning impact.

- **The Technical Complexity Dimension.** The technical complexity refers to the need for more or less complex technologies in a simulation. For instance, a 2D virtual environment with mouse movements and keyboard strokes as actions for the user's avatar may be good enough for certain pedagogical objectives. For other objectives, more complex modules such as force feedback or haptic devices, touch screens, 3D sound and object models, augmented/mixed reality features, may be justifiable. When designing an RVS, one should limit the technical complexity of the simulation interface to only what is needed to achieve functional task alignment towards a satisfying learning impact.

**DEFINITIONS OF RVS AND RVS ECOSYSTEM**

To sum up the previous section, an RVS is defined as a techno-pedagogical activity that is designed following a realistic-enough philosophy, i.e. whose aim is to achieve a satisfying learning impact.

However, one will maximize the benefits of a given RVS if it is associated with other technological modules to form an *RVS Ecosystem* (RVSE) that supports rapid adaptability to learners' and educators' needs. Consequently, an RVSE is defined by four rapid-related objectives:

i. **Rapid creation.** Tools called Editor applications must be developed following an expert-centered design. This permits rapid and autonomous creation of RVS by scaffolding non-technical expert actions to enforce a realistic-enough philosophy.

ii. **Rapid deployment and scalability.** An RVSE must contain technological modules that allow RVS contents to be easily shared to many learners anytime anywhere.

iii. **Rapid data-driven feedback.** An RVSE must use data generated by learner activities in order to facilitate (a) pedagogical interactions between learners and teachers (e.g. debriefing), and (b) tutor rapid overview of RVS performances of a cohort of learners.

iv. **Rapid trouble detection**. An RVSE should detect and rapidly inform creators of problems within RVS. Fixes and updates can then be facilitated by editor applications discussed above.

## SUGGESTED MODULAR ARCHITECTURE FOR RAPID VIRTUAL SIMULATION ECOSYSTEMS

Figure 1 presents a suggested modular RVSE architecture. First we define and give examples of an RVS Editor Application and an RVS Client Application, two foundational components of an RVSE. Then we describe how these two modules, along with other ones, work together to satisfy the 4 rapid-related objectives listed earlier.

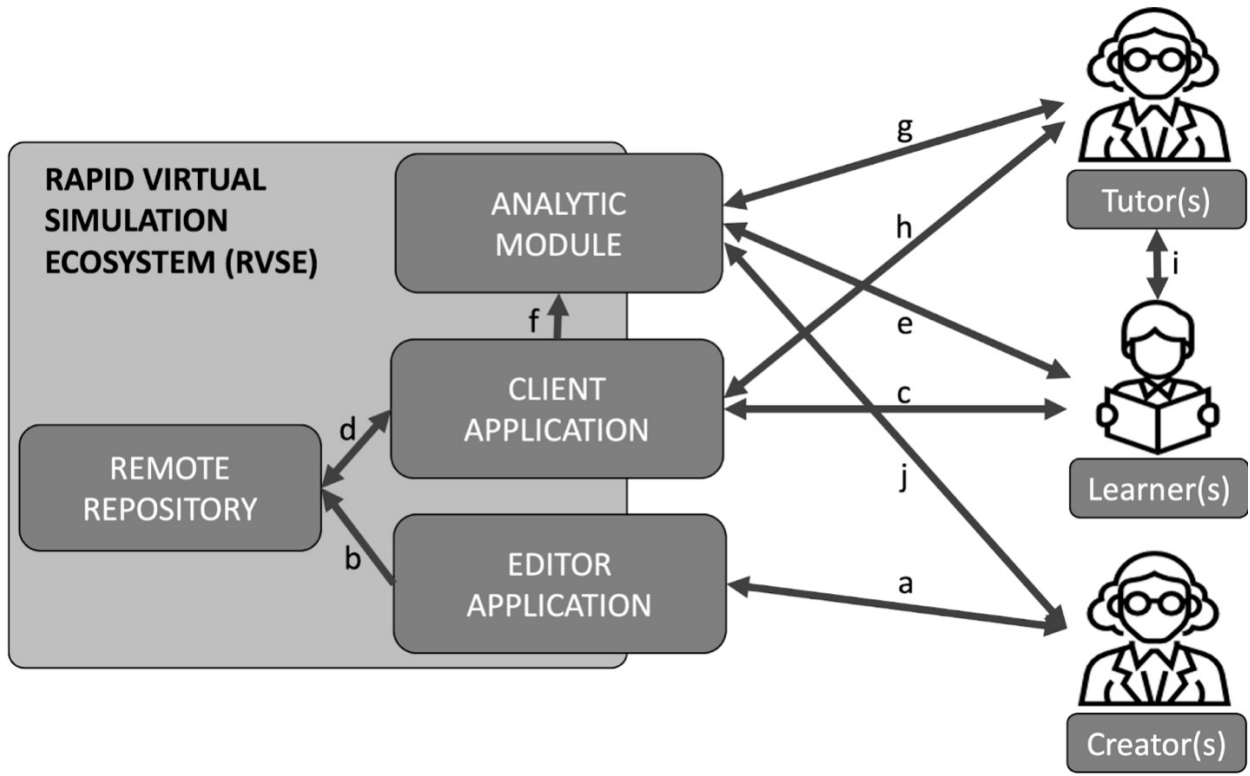

Figure 1. A suggested RVSE Architecture.

**RVSE Foundational Components**

The RVS Editor Application and the RVS Client Application are the principal interfaces encountered by users of an RVSE. An RVS Editor Application is used to create RVS activities such as patient case simulations whereas the RVS Client Application is used to present them to learners.

The Deteriorating Patient App project (DPApp)[36] is a health sciences RVS that simulates a "patient" whose portrayed clinical states deteriorate over 10 minutes unless learners select priority actions to take in "diagnosing" and "stabilizing" it. Figures 2 and 3 are screenshots of the Editor and Client applications from the DPApp.

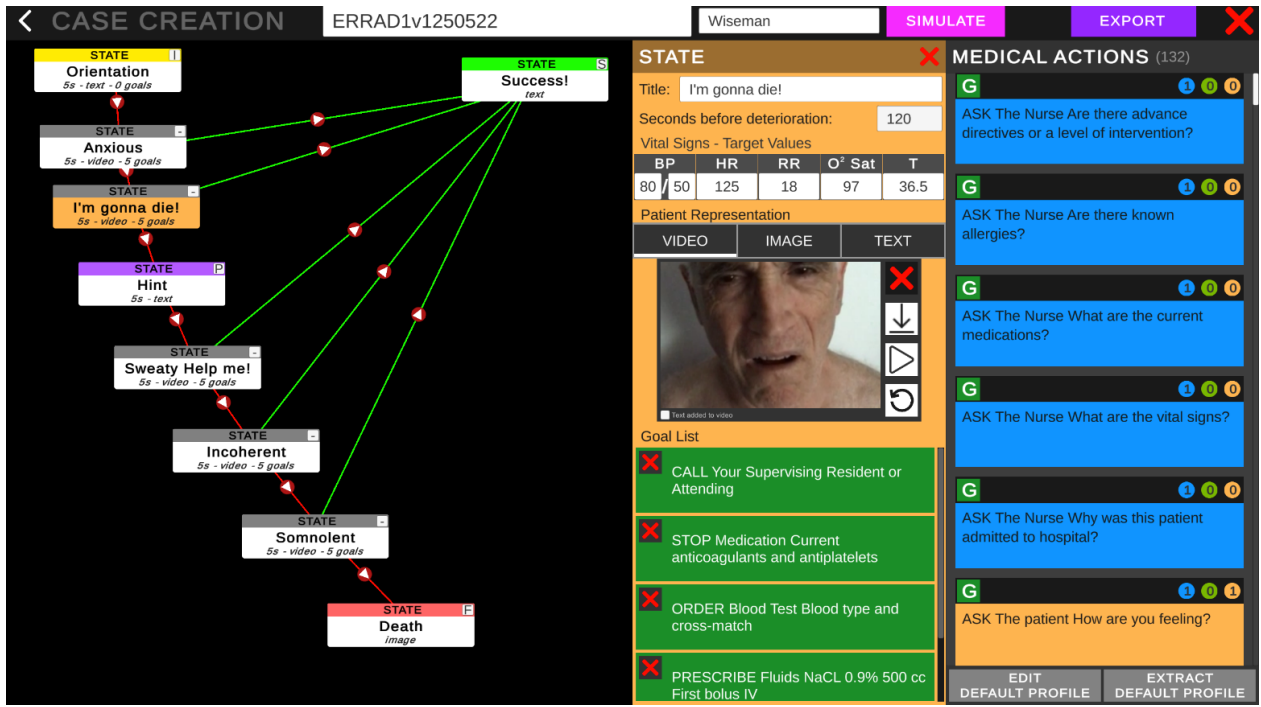

Figure 2. Screenshot of the DPApp Editor Application.

The DPApp Editor Application contains three panels that are used to scaffold case creation. In the leftmost panel, a graph (or flow diagram) models how the patient deteriorates over time. The center panel displays information (visual representations, vital signs, goal actions, duration) that define each state of the graph (i.e. steps in the flow diagram), and the rightmost panel displays a set of all the possible medical actions to perform on the simulated patient.

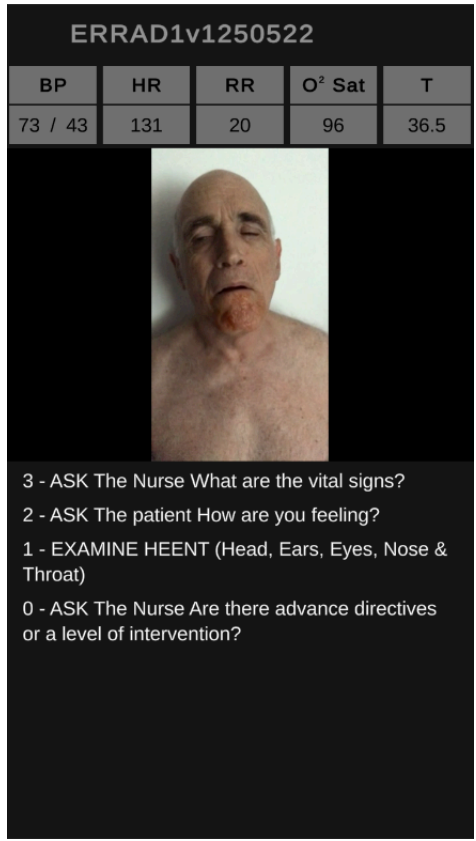

Figure 3. Screenshots of the DPApp Client Application.

A simulated patient created with the DPApp Editor Application is presented to learners by the DPApp Client Application. The current state of the simulated patient is portrayed by a *top bar* showing evolving vital signs and a *central area* displaying a patient representation that could be a series of text descriptions, photos or short videos. Clicking on this central area allows learners to select a medical action to perform on the simulated patient. Finally, the lower part lists all actions previously performed by the learner during the activity.

**Enacting the Rapid-Related Objectives of an RVSE**

Following are explanations on how each rapid-related objective can be achieved in the suggested architecture presented in the previously presented figure 1:

- **Rapid Creation.** An expert with a creator role uses an RVS Editor Application (a) to produce an RVS activity.
- **Rapid Deployment and Scalability.** Once the produced RVS activity is deemed finalized, it is uploaded into a Remote Repository (b). This storage module is located on the internet (or on a private network). When learners want to do an RVS activity or are requested to do it by a tutor (i), they launch an RVS Client application (c) and select the appropriate content from an embedded catalog of available activities obtained from the Remote Repository (d). Tutors can also access the client to preview available RVS activities before including them in their teaching sessions (h).
- **Rapid Data-Driven Feedback**. RVS activities produce data that are transmitted to, and processed within an RVS Analytic Module (f) to create a variety of dashboards whose goal is to make individual and group performances easily visible to, and analyzable by learners and educators. For instance, dashboards could display cohort performance to tutors (g) or individual performances to learners (e), or support tutors organizing in-person or online debriefing activities with learners (g-i). Such data-driven feedback could support synchronous as well as asynchronous activities, the latter making education possible at a much larger scale.
- **Rapid Trouble Detection.** RVS Analytic modules are equipped with data-driven AI algorithms detecting problems within RVS activities (f), and triggering alarm messages sent to creators (j). There are at least 3 possible explanations for such problems. Firstly, the RVS activity may be ill designed, in which case the creator has to provide corrections with the help of the RVS Editor Application (a). Secondly, there may be a dissonance between the current level of learners, and challenges posed by RSV activities. Thirdly, concepts and skills required to solve RVS activities may not have been adequately taught.

**ADVANTAGES AND CHALLENGES OF RVSE-BASED EDUCATION**

As a novel techno-pedagogical tool centered on the Realistic-Enough Philosophy, an RVSE brings several advantages:

- It is repeatable, space and time independent, scalable, and allows for data-driven smart assessment (see introduction) as well as spaced learning.
- It liberates manpower, space, and equipment resources, which then become available for other educational activities.
- It reduces the time and cost needed for creating, correcting, or improving simulation cases.
- It simplifies simulation case creation by domain experts, thereby increasing their autonomy and reducing their dependency on external professionals.
- It records and make performance data readily visible to both learners and teachers to guide debriefing activities.

**RVSE-based education also presents several challenges:**

- It demands early clarification of intended learning impacts, along with criteria to categorize them as 'satisficing'.
- It requires Client application mechanics and interfaces that are easy and quick to master so that learners spend more time on educational activities than on learning how the technology works.
- Creating an effective editor application requires several types of expertise:
    - mastery of the expert domain,
    - a clear understanding of the pedagogical advantages and disadvantages of various technical choices, as well as the complexity of implementing them,
    - a good knowledge of the characteristics of target users (e.g.: level of expertise, technological literacy).

- Unsupervised and easy simulation access inherent to RVSE-based education carries a risk of trivialization on the part of learners, who could see them as games to play and win rather than opportunities to learn.
- It requires other complementary educational activities such as prebriefing, debriefing, and feedback in order to be fully effective.

**REASONS FOR AND AGAINST THE ADOPTION OF RVSE-BASED EDUCATION**

There are several reasons for the adoption of RVSE-base education:
- **To augment learning opportunities.** RVSE-based education provides learners with interactive learning activities that require few resources and little time to experience, anytime, anywhere. This flexibility makes it possible for very busy students and educators to participate in learning activities more frequently.
- **To facilitate blended learning.** An RVSE allows for the creation of virtual simulations tailored to a blended sequence of virtual and in-person educational activities. For example, the sequence can begin with a virtual RVS-based simulation to introduce general principles (e.g. setting priorities in stabilizing a patient with hemorrhagic shock). Subsequently, students would come to a high fidelity mannequin in a full service center to apply learned principles and practice relevant skills (e.g. working as part of a team, obtaining urgent intravenous access). Finally, students could be offered additional RVS-based simulations as booster educational activities for reinforcement learning anytime anywhere (e.g. learning to use the correct site and caliber of intravenous access and appropriate intravenous therapy).
- **To optimize expert involvement**. Many health science education experts are active clinical practitioners whose busy schedules leave scant time for educational practices. In RVSE-based education, the Editor application helps experts efficiently convert their attitudes, skills and knowledge into simulation activities whereas the Client application allows these simulation activities to be easily disseminated. This reduces experts' space and time constraints, freeing them for other clinical and educational obligations.

- **To improve educational practices**. RVSE-based education requires educators to adjust learning tasks to learners' current level of attitudes, skills and knowledge, and to strive to make the "implicit explicit"[37] when creating learning tasks. This is conducive to self reflection on one's educational practices, and fosters further development of educational competencies. Furthermore, feedback provided to educators by data-driven analytics and visualizations can help them refine their teaching approaches.

- **To improve educational agility.** RVSE-based education supports agile educational development: inspired by a software engineering approach, the concept of agility in RVSE-based education implies the creation and modification of simulations via rapid cyclic iterations. The resulting simulations are better designed, more robust, more quickly adaptable, and more easily distributed in response to changing contextual conditions and crises such as COVID pandemics.

- **To optimize institutional resource management.** Because RVSE-based education is scalable, repeatable, and once created, requires limited technological capabilities and materials. It thereby optimizes institutional resource management by simplifying and reducing the cost of human resources (experts and support personnel), space, high fidelity simulation materials, and overall logistics.

However there are also situations where RVSE-based education may not be suitable:

- **Satisficing options already exist**. RVSE-based education is unnecessary If there are already available educational methods or techno-pedagogical tools that are familiar for the community of users and experts, cheap enough and good enough at fulfilling a given learning objective.

- **There is a lack of design expertise or of preexisting similar solutions.** RVSE-based education will be hard to set up and deploy if there is neither an available team with the necessary interdisciplinary expertise to create it, nor a similar RVSE that could be adapted to address newly targeted learning objectives.
- **The technological infrastructure is inadequate.** RVSE-based education requires limited technological capabilities and materials. However, it still requires basic and reliable technological infrastructure such as electricity, access to electronic devices (smartphones, computers), and internet access in some cases.

**CONCLUSION**

In this article we have presented the concept of *Rapid Virtual Simulations*, a new techno-pedagogical activity that is grounded in a *realistic-enough philosophy* for simulation-based education, and which strives to achieve a *satisficing learning impact*. We have also presented the concept of a *Rapid Virtual Simulation Ecosystem* as an integrated set of technological modules designed to facilitate experts' engagement and tasks by considering their needs and constraints, and to multiply educational affordances for learners. As an example, we have shown how several ecosystem modules have been implemented in a multiyear interdisciplinary project called the Deteriorating Patient App. We have discussed the pros and cons of Rapid Virtual Simulation Ecosystems to identify situations where they will be particularly beneficial to educational systems, as well as situations that limit their usefulness.

With Rapid Virtual Simulations and their associated Ecosystems, we argue for technological agility and simplicity as key guiding principles for the design of future educational systems. Rapid Virtual Simulations, as scalable, fast, frugal and easily-adjustable learning activities, have the potential to provide efficient education of new skills to improve societal resilience during future crises. This is an important focus of our future work.

Finally, we wonder how additional ecosystem modules could foster collaboration between global healthcare educational programs and institutions as co-creators of interactive pedagogical resources, and as benevolent collaborators rather than as adversaries interested in intellectual property, reputation, and academic domination.